\documentclass[twocolumn,preprintnumbers,superscriptaddress,nofootinbib,aps,prd,floatfix]{revtex4}

\usepackage{amsmath,amssymb}
\usepackage{graphicx} %loads the graphicx.sty package 
\usepackage{epstopdf} %loads the epstopdf.sty package 
\usepackage{slashed}
\usepackage{color}
\usepackage{multirow}
\usepackage{hyperref}
\usepackage[utf8]{inputenc}

\usepackage{footmisc}

\usepackage{array}
\newcolumntype{L}[1]{>{\raggedright\let\newline\\\arraybackslash\hspace{0pt}}m{#1}}
\newcolumntype{C}[1]{>{\centering\let\newline\\\arraybackslash\hspace{0pt}}m{#1}}
\newcolumntype{R}[1]{>{\raggedleft\let\newline\\\arraybackslash\hspace{0pt}}m{#1}}

\newcommand{\be}{\begin{eqnarray*}}
\newcommand{\ee}{\end{eqnarray*}}

\newcommand{\bee}{\begin{eqnarray}}
\newcommand{\eee}{\end{eqnarray}}
\newcommand{\beeq}{\begin{equation}}
\newcommand{\eeeq}{\end{equation}}

\usepackage{xspace}

\begin{document}

\title{{Mapping the shape of the scalar potential with gravitational waves}}

\begin{abstract}
\end{abstract}
\author{Mikael Chala} %
\author{Valentin V. Khoze} %
\author{Michael Spannowsky} %
\author{Philip Waite} %

\affiliation{Institute for Particle Physics Phenomenology, Department
  of Physics,\\Durham University, DH1 3LE, United Kingdom\\[0.1cm]}
  
\preprint{IPPP/19/34}

\begin{abstract}
\noindent We study the dependence of the observable stochastic gravitational wave background induced 
by a first-order phase transition on the global properties of the scalar effective 
potential in particle physics. The scalar potential can be that of the Standard Model Higgs field, or more generally
of any scalar field responsible for a spontaneous symmetry breaking in beyond-the-Standard-Model settings that
provide for a first-order phase transition in the early universe.
Characteristics of the effective potential include the relative depth of the true minimum ($E_\alpha^4$), the 
height of the 
barrier that separates it from the false one ($E_m^4$) and the separation between the two minima in field space ($v$), 
all at the bubble nucleation temperature. We focus on a 
simple yet quite general class of single-field polynomial potentials, with 
parameters being varied over several orders of magnitude. It is then shown that 
gravitational wave observatories such as aLIGO O5, BBO, DECIGO and LISA are 
mostly sensitive to values of these parameters in the region $E_\alpha 
\sim (0.1-10) \times E_m$. Finally, relying on well-defined models and using our framework, we 
demonstrate how to obtain the gravitational wave spectra for potentials of various shapes  
without necessarily relying on dedicated software packages.
\end{abstract}

\maketitle

\section{Introduction}
\label{sec:intro}
The first detection of gravitational waves (GW) on Earth by the LIGO 
collaboration in 2016~\cite{Abbott:2016blz} opened a new window to explore 
high-energy 
physics phenomena. One such source of gravitational radiation are 
first-order phase transitions (FOPT), which occur when a scalar field tunnels from 
a local minimum to a lower-lying true vacuum that is separated by an energy 
barrier~\cite{Quiros:1999jp}. %

FOPTs proceed via the nucleation 
of bubbles of the stable true vacuum in the 
meta-stable false vacuum phase. The phase transition occurs at the temperature
$T=T_*$  where bubbles of critical size can be formed; these critical bubbles 
expand, collide and ultimately thermalise by releasing their latent heat energy into the 
plasma formed
of light particles.

The main frequency of the corresponding stochastic GW 
background grows with $T_*$.~(Future experiments targeted at growing values of 
$T_*$ include LISA, BBO, DECIGO or 
aLIGO O5; see Ref.~\cite{Moore:2014lga} for details.)~However, it is 
not yet clear how this frequency as well as the corresponding amplitude depend 
on the global properties of the scalar potential. We address this question in 
this paper.

To this aim, we focus on a class of polynomial functions parametrised by
\begin{equation}\label{eq:parametrisation}
 V_{T_*}(\varphi) = \left[\left(\frac{\varphi}{c}\right)^2-a\right]^2 + b 
\left(\frac{\varphi}{c}\right)^3~,
\end{equation}
describing the shape of the scalar potential density evaluated at the 
temperature $T_*$ where the phase transition happens. Let us 
emphasize that this functional form is merely a useful proxy that allows 
us to (numerically) trade the parameters
$a$, $b$ and $c$ for the values of the vacuum expectation value (VEV) of 
$\varphi$ in the true minimum $v$, its depth $(E_\alpha^4)$ and the energy 
barrier $(E_m^4)$. Almost \textit{every potential} can be well characterised by 
these parameters, as we demonstrate in subsequent sections; therefore our study 
does not restrict to Eq.~\eqref{eq:parametrisation} by any means. 

\begin{figure}[!t]
 \includegraphics[width=\columnwidth]{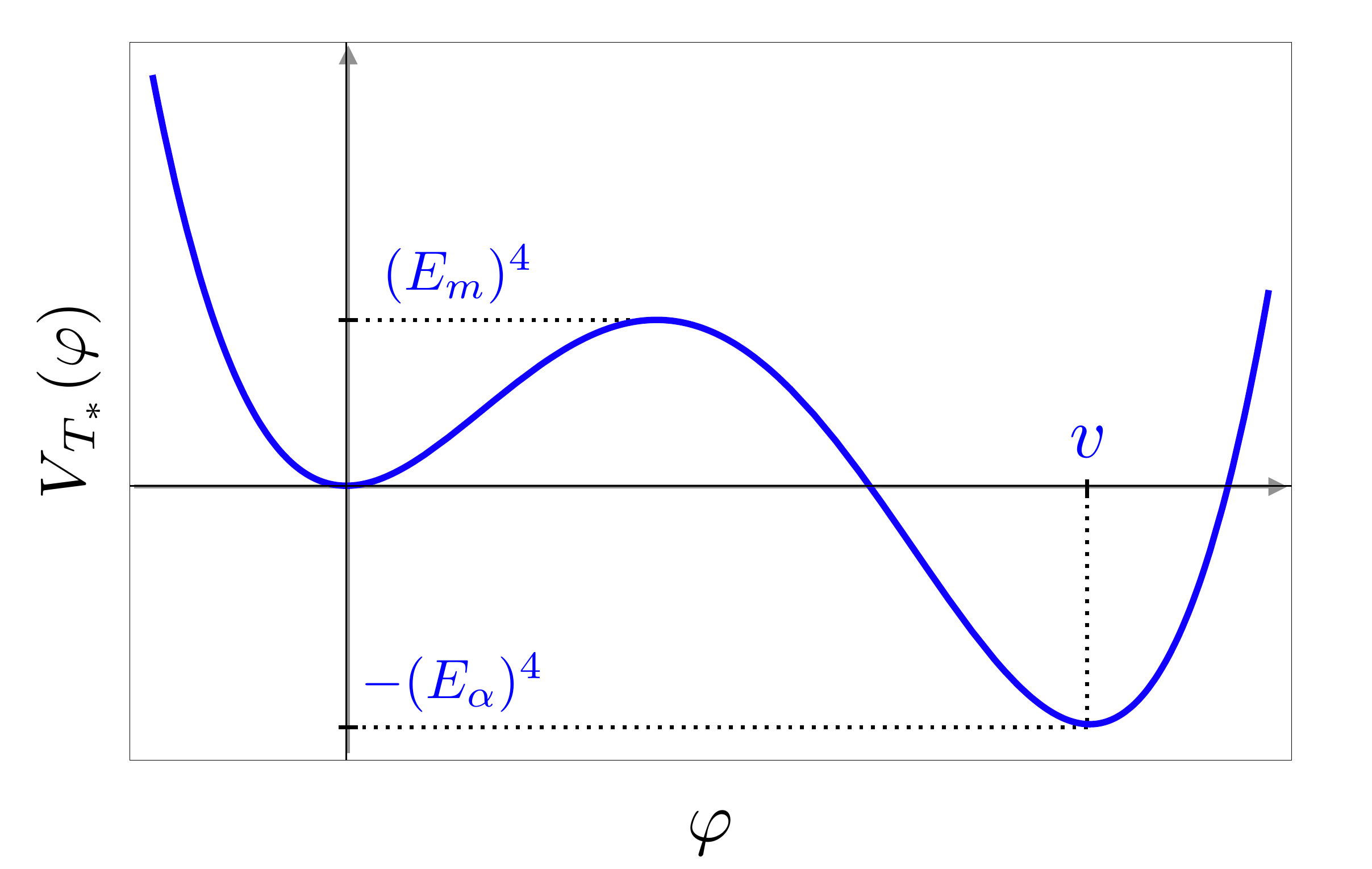}
 \caption{\it The scalar potential of a generic particle physics model with a 
FOPT. 
 The potential is computed at the nucleation temperature $T_*$ where the 
nucleation rate $P$ to form bubbles of the true vacuum 
 approaches $P\simeq 1$.}\label{fig:pot1}
\end{figure}

The field $\varphi$ 
can be 
the Standard Model 
Higgs~\cite{Zhang:1992fs,Grojean:2004xa,Delaunay:2007wb,Grinstein:2008qi,
Harman:2015gif,Chala:2018ari} or more 
generally any new other scalar 
field~\cite{Schwaller:2015tja,Jaeckel:2016jlh,Addazi:2017gpt,Breitbach:2018ddu,
Croon:2018erz,Croon:2019iuh,Dev:2019njv}.
Without loss of generality, the potential is finally shifted 
in order for the 
false minimum to lie at the origin; see Fig.~\ref{fig:pot1}.

Because the main aim of this article is understanding how 
$T_*$ as well as other quantities relevant for the computation of 
the GW stochastic background depend on $v$, $E_m$ and $E_\alpha$, we compute 
the former parameters
varying the later over several orders of magnitude. 
 This procedure is explained in 
detail in 
Section~\ref{sec:con}. For numerical calculations in this Section we rely 
predominantly on \texttt{CosmoTransitions}~\cite{Wainwright:2011kj} and 
\texttt{BubbleProfiler}~\cite{Athron:2019nbd} and cross-check these tools using 
the neural 
network method introduced in Ref.~\cite{Piscopo:2019txs}\footnote{We acknowledge that 
various other methods exist to calculate the bubble profiles or tunnelling rates 
\cite{John:1998ip,Konstandin:2006nd,Masoumi:2016wot,Akula:2016gpl,Jinno:2018dek,
Espinosa:2018hue,Espinosa:2018szu,Guada:2018jek}.}. 

We compute the actual GW signal in Section~\ref{sec:gw}, and discuss its 
dependence on the global properties of the potential. In Section~\ref{sec:ligo} 
we estimate the reach of different GW facilities, including LISA, DECIGO, BBO 
and aLIGO O5. 

In Section~\ref{sec:models}, we explain how to use our results to compute the 
GW spectrum in well-defined models of new physics. We validate this method by 
comparing to exact numerical integration of the action in each model.
Finally, 
we offer conclusions in Section~\ref{sec:conclusion}.

\section{Parametrisation of the effective potential}
\label{sec:con}

Our starting point is the effective potential $V_{T_*}(\varphi)$ in Eq.~\eqref{eq:parametrisation} that corresponds to a general particle physics model 
at the temperature $T=T_*$,
where the model undergoes a FOPT. $T_*$  is the temperature of the formation of critical bubbles
and is usually referred to as the nucleation temperature. 
%
% %
In the unbroken phase, the VEV of $\varphi$ is 
vanishing,
$\langle\varphi\rangle=0$, while in the broken phase it is non-zero, 
$\langle\varphi\rangle=v$.

Without loss of generality, we assume that the vacuum at the origin is the 
false minimum; the vacuum with the non-zero VEV being the true global one 
with
vacuum energy $V_{T_*}(v)=-(E_\alpha)^4<0.$ 
In total, the effective potential in Fig.~\ref{fig:pot1} is characterised by 
three real-valued and positive parameters of 
mass-dimension one: the vacuum separation $v$, the vacuum energy change parameter $E_\alpha$, and the 
barrier height parameter $E_m$. 

The value of the nucleation temperature is determined from the 
requirement that the probability ($P$) for a single bubble to nucleate within 
the horizon volume is of order one~\cite{Moreno:1998bq}:
\begin{equation}
P(T_*) \,=\, \int_\infty^{T_*} \frac{dT}{T}\,\left(\frac{2 \zeta M_{\rm Pl}}{T}\right)^4 \exp\left[ -\frac{1}{T}\, S_3^{\,\rm cl}(T)\right]\simeq 1
 \,,
\label{eq:P1}
\end{equation}
%
% \begin{equation}
% %T_* :\,\, 
% P(T_*) \simeq 1 \quad \Rightarrow \quad \frac{1}{T_*}\, S_3^{\,\rm cl}(T_*) \,\simeq\, 100\,,
% \label{eq:100}
% \end{equation}
%
where $S_3^{\, \rm cl}(T_*)$ is the action 
computed on the classical $O(3)$-symmetric bounce solution\footnote{We have checked that the $O(4)$-symmetric bounce solution has generally a much larger action (in agreement with the claim often made in the literature~\cite{Quiros:1999jp,Caprini:2015zlo} that it is only relevant for vacuum transitions). This fails only in points with $T/v \ll 1$ which, as we discuss further in next sections, are physically questionable. We therefore restrict to the $O(3)$-symmetric bounce.} in the 
3-dimensional theory with the potential 
$V_{T_*}(\varphi)$~\cite{Coleman:1977py,Linde:1981zj}.
% The second estimate in Eq.~\eqref{eq:100} follows from the expression for the 
% bubble nucleation rate,
%see e.g. Ref.~\cite{Moreno:1998bq}:
%
% \begin{equation}
% P(T_*) \,=\, \int_\infty^{T_*} \frac{dT}{T}\,\left(\frac{2 \zeta M_{\rm Pl}}{T}\right)^4 \exp\left[ -\frac{1}{T}\, S_3^{\,\rm cl}(T)\right]
%  \,,
% \label{eq:P1}
% \end{equation}
%where 
We have also defined $\zeta^{-1}=4 \pi \sqrt{\pi g_*(T)/45}$. For the effective number of relativistic degrees of freedom in the plasma
$g_*(T_*)  \sim 100$, we have $\zeta = 0.03$.
 To allow the expression on the right-hand side of Eq.~\eqref{eq:P1} to be of 
order one,
the exponential suppression factor should be compensated by the large prefactor 
in Eq.~\eqref{eq:P1}:
\begin{align}\nonumber
\frac{1}{T_*} S_3^{\,\rm cl}(T_*) &\,\simeq 4 \log\left(0.06  \frac{M_{\rm 
Pl}}{T_*}\right)\\
&\,\simeq140 -4 \log \frac{T_*}{100 \,{\rm GeV}}\,.\label{eq:100}
\end{align}
%
% which we approximate as 100, thus confirming the estimate quoted above in 
% Eq.~\eqref{eq:100}.
For FOPTs at $T_*\sim 100$ GeV, the second term in the last equation can be neglected, leading to the usual approximation $S_3^{\,\rm cl}(T_*)/T_*\sim 140$. We are however interested in FOPTs at arbitrarily large $T_*$, so we will take the full temperature dependence into account in what follows.

To optimise the scanning procedure over effective potentials with different global properties it is useful to introduce 
dimensionless variables by rescaling all physical parameters of the potential 
in Fig.~\ref{fig:pot1} with respect to 
a single overall scale. A convenient choice for our purposes is the VEV $v$ of 
the global 
minimum\footnote{Note 
that
the effective potential and all its parameters are defined at the fixed value 
of $T=T_*$.
Hence the quantities in Eq.~\eqref{eq:tildes} are $v=v(T_*)$, 
$E_\alpha=E_\alpha(T_*)$ and $E_m=E_m(T_*)$.}.

\begin{figure}[!t]
 \includegraphics[width=\columnwidth]{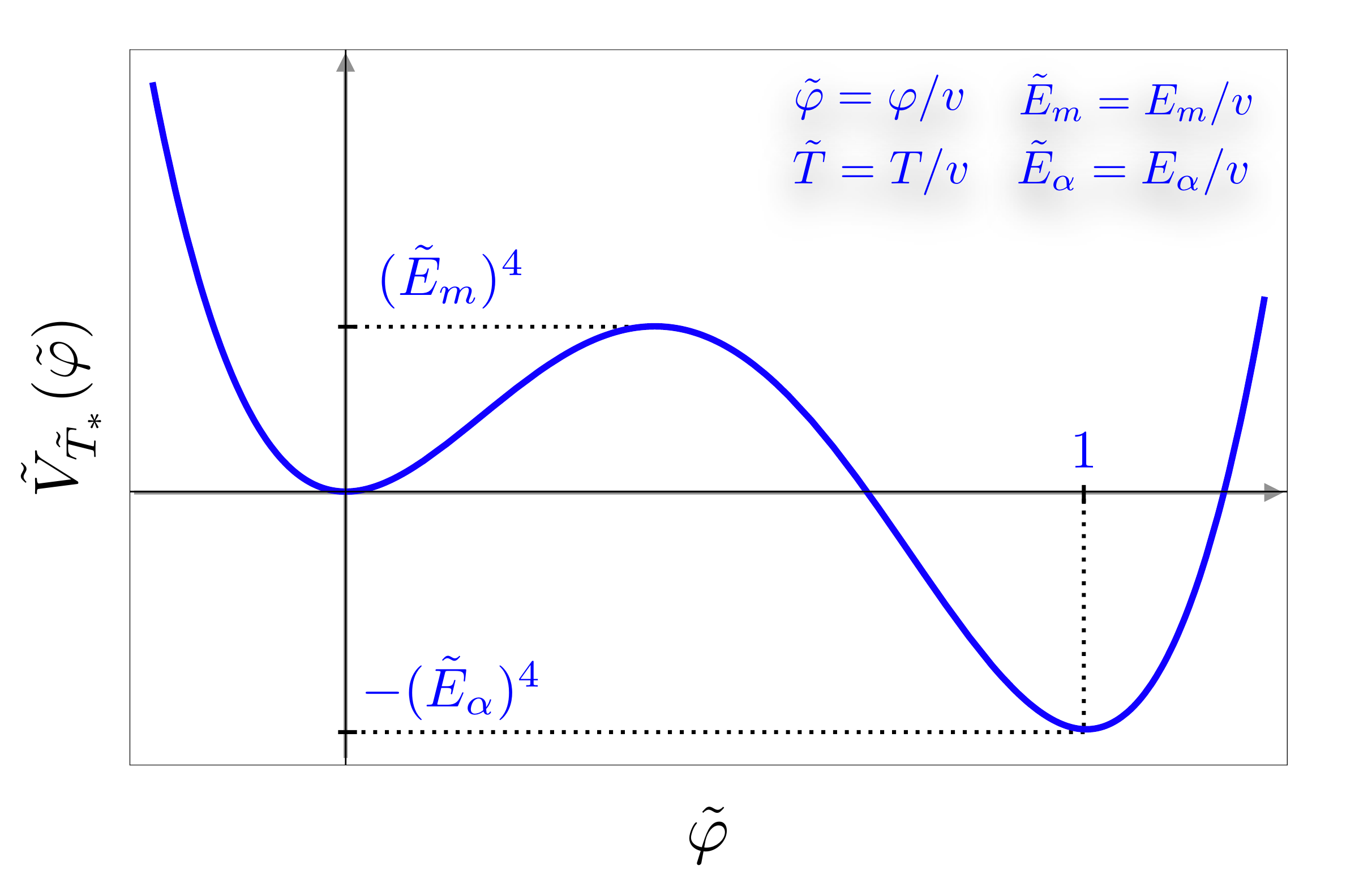}
 \caption{\it The effective potential in the rescaled variables.}\label{fig:pot2}
\end{figure}

We define:
\begin{eqnarray}
\tilde{\varphi}(x)=\frac{\varphi(x)}{v}~, \,\, \tilde{T}= \frac{T}{v}~, \,\,
%\nonumber\\
\tilde{E}_\alpha= \frac{E_\alpha}{v}~, \,\, \tilde{E}_m= \frac{E_m}{v}~.
\label{eq:tildes}
\end{eqnarray}
Upon rescaling with $v$, the corresponding %
potential $\tilde{V}_{\tilde{T}_*}(\tilde{\varphi})$ is shown in 
Fig.~\ref{fig:pot2}
and is characterised
now by two free parameters, $\tilde{E}_\alpha$ and $ \tilde{E}_m$, with the minima fixed at $\tilde{\varphi}=0$ and 
$\tilde{\varphi}=1$.

For any given effective potential at the nucleation temperature $\tilde{T}_*$, we can now compute the
value of  $\tilde{T}_*$ using Eq.~\eqref{eq:100}. To this end, we first need to 
find the $O(3)$-symmetric classical bounce solution 
that extremises the Euclidean action of the 3-dimensional theory with the 
potential $\tilde{V}_{\tilde{T}_*}(\tilde{\varphi})$,
\begin{equation}
\tilde{S}_3 \,=\, 4\pi \int dr\, r^2 \left(\frac{1}{2} \tilde{\varphi}^{\, \prime}(r)^2 + \tilde{V}_{\tilde{T}_*}(\tilde{\varphi})
 \right)\,,
 \label{eq:S3d}
\end{equation}
by solving the classical equation~\cite{Coleman:1977th},
\begin{equation}
\tilde{\varphi}^{\,\prime\prime} (r) + \frac{2}{r} \tilde{\varphi}^{\, \prime}(r) \,=\,
 \partial_{\tilde{\varphi}} \tilde{V}_{\tilde{T}_*}(\tilde{\varphi})\,.
\label{eq:class}
\end{equation}
We use custom routines based on \texttt{BubbleProfiler}~\cite{Athron:2019nbd} to 
this 
aim. We subsequently compute the action on this classical bounce solution, 
$\tilde{S}_3^{\,\rm cl}$, and finally impose the bound of Eq.~\eqref{eq:100}
to find
%
% \begin{equation}
%  \tilde{T}_* \,=\, %\frac
%  \frac{{\tilde{S}_3^{\,\rm cl}}}{{100}}\,, \quad {\rm or} \quad {T}_* \,=\, 
% v\,%\frac
%  \frac{{\tilde{S}_3^{\,\rm cl}}}{{100}}\,.
%  \label{eq:Tc}
% \end{equation}
%
\begin{equation}
\frac{{\tilde{S}_3^{\,\rm cl}}}{\tilde{T_*}} \simeq 140 - 4 \log \frac{\tilde{T}_*}{100} -4\log\frac{v}{\text{GeV}}\,.
 \label{eq:Tc}
\end{equation}
We determine the nucleation temperature $T_*$ by solving (numerically) Eq.~\eqref{eq:Tc}\footnote{To this aim, we fix $v=100$ GeV, although we note that the parameter $\log (v/\text{GeV})$ does not correct our result in Eq.~\eqref{eq:Tc} by more than $20\%$ unless $v$ is very large, $v\gg 10^{6}$ GeV.}. This
is the first of the three main parameters we need to obtain the 
stochastic GW spectrum generated in the FOPT.

The second parameter affecting the GW spectrum is the latent heat $\alpha$. 
It is defined as the ratio of the
energy density released in the phase transition to the energy density of the
radiation bath in the plasma:
\begin{equation}
\alpha =\, \frac{\rho_{\rm vac}}{\rho_{\rm rad}} = \frac{{E}_\alpha^4}{g_*(T_*)\, \pi^2\, T_*^4/30}\,\simeq\,
0.03 \, \left(\frac{\tilde{E}_\alpha}{\tilde{T}_*}\right)^4.
\label{eq:alph}
\end{equation}

The third quantity we need is $\beta/H_*$, characterising the speed of 
the phase transition:
\begin{equation}
\frac{\beta}{H_*}\, =\, T_*\frac{d}{dT}\left(\frac{1}{T}\,S_3^{\,\rm 
cl}(T)\right)_{T=T_*}.
\label{eq:beta}
\end{equation}
In this equation, $H_*$ represents the Hubble constant at the time when 
$T=T_*$.
%, .
A strong GW signal results from a slow phase transition with a large latent 
heat release, i.e. in the small $\beta/H_{*}$ 
and large $\alpha$ regime.

\begin{figure}[!t]
 \includegraphics[width=\columnwidth]{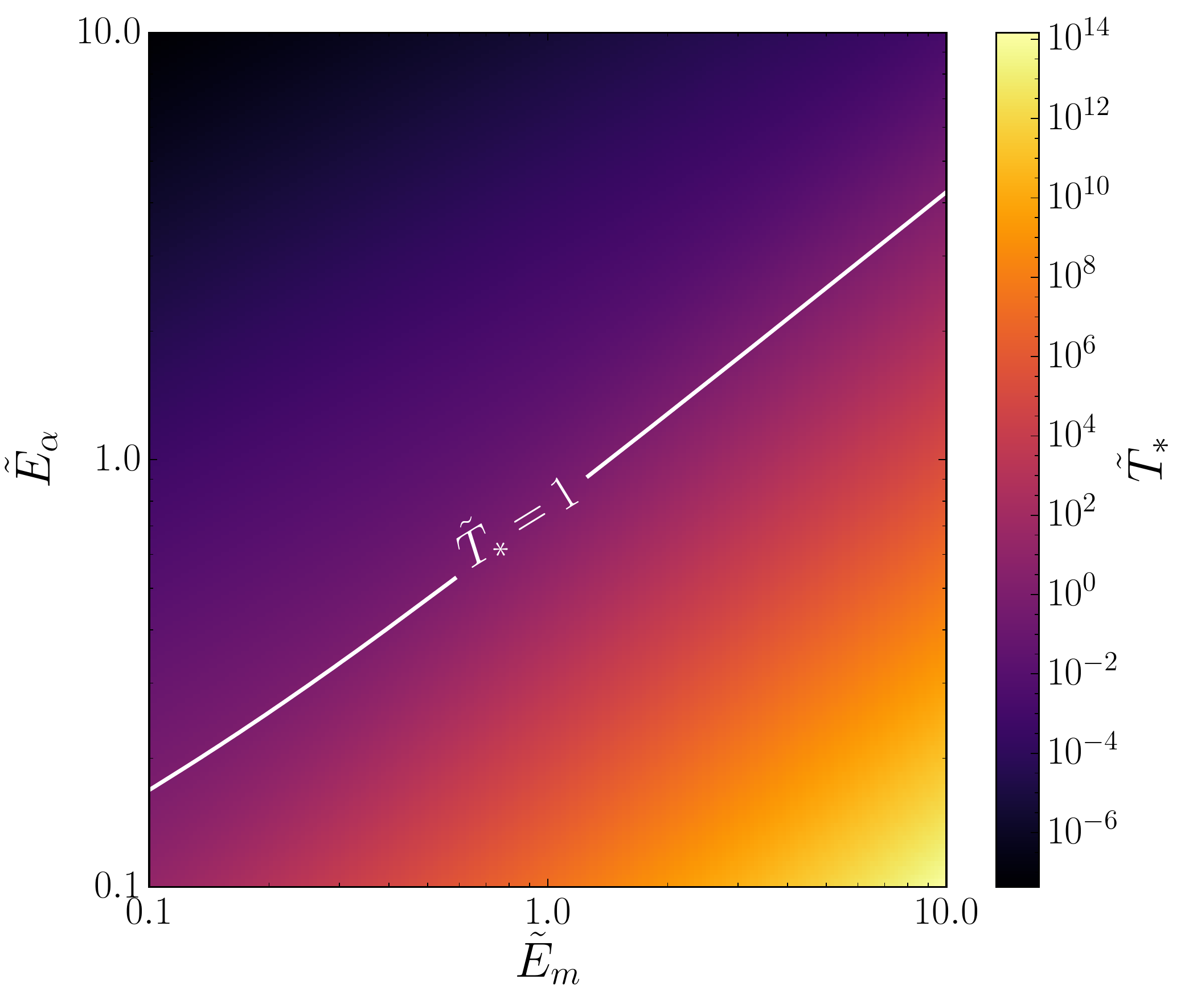}
 \caption{\it Values of $\tilde{T}_*$ as a function of $\tilde{E}_m$ and 
$\tilde{E}_\alpha$ in the range $[0.1, 10.0]$. Note that since the potential density $V_T(\varphi)$ depends 
on the fourth power of $\tilde{E}_m$ and $\tilde{E}_\alpha$, the resulting variation of the shape of the potential is over 
eight orders of magnitude.}\label{fig:temperatures}
\end{figure}

To determine  $\beta/H_*$ from Eq.~\eqref{eq:beta}, we need to know the slope 
of the classical action $S_3^{\,\rm cl}(T)$ at $T=T_*$, 
and hence we need to compute infinitesimal deviations of the effective potential $V_T(\varphi)$ from its value 
at the nucleation temperature. One could use the full temperature-dependent 
expression for the effective potential,
%
%$
at 1-loop level \cite{Dolan:1973qd},
%
%\begin{widetext}
\begin{align}\label{VTgen}
\Delta V_T &= V_T -V_{T=0} \\
&= \frac{T^{4}}{2\pi^{2}}\sum_{i}\pm 
n_{i}\int_{0}^{\infty}\mbox{d}q\, 
q^{2}\log\bigg[1\mp e^{-\sqrt{q^{2}+m_{i}^{2}(\varphi)/T^{2}}}\bigg],\nonumber
\end{align}
%\end{widetext}
%
but this approach would require us to specify the details of the mass spectrum $m_i(\varphi)$
and of the number of degrees of freedom $n_i$  in the microscopic theory. To retain a large degree of model-independence
for our considerations, we use instead the leading-order Taylor expansion 
approximation, which is fully justified at 
high temperatures $T_* > \varphi$:
\begin{equation}
V_T (\varphi)\,=\, V_{T_*} (\varphi) \,+\, a_T\, (T^2-T_*^2)\, \varphi^2\,.
\label{VTa}
\end{equation}

There is just a single new parameter $a_T$ on the right-hand side of 
Eq.~\eqref{VTa} that incorporates all model-dependence and 
characterises the deviations of $T$ from $T_*$ for different models. For any 
specific model the value of $a_T$ can be obtained 
upon expanding Eq.~\eqref{VTgen} to the order $T^2 \varphi^2$ in the 
high-temperature expansion, $m_i(\varphi)^2/T_*^2 <1$. This gives:
\begin{equation}
a_T\,=\,  \frac{1}{24}\,\sum_{b,f} (n_{b} +n_f/2) \,,
\end{equation}
where the sum is over bosonic and fermionic degrees of freedom.

\begin{figure}[!t]
 \includegraphics[width=\columnwidth]{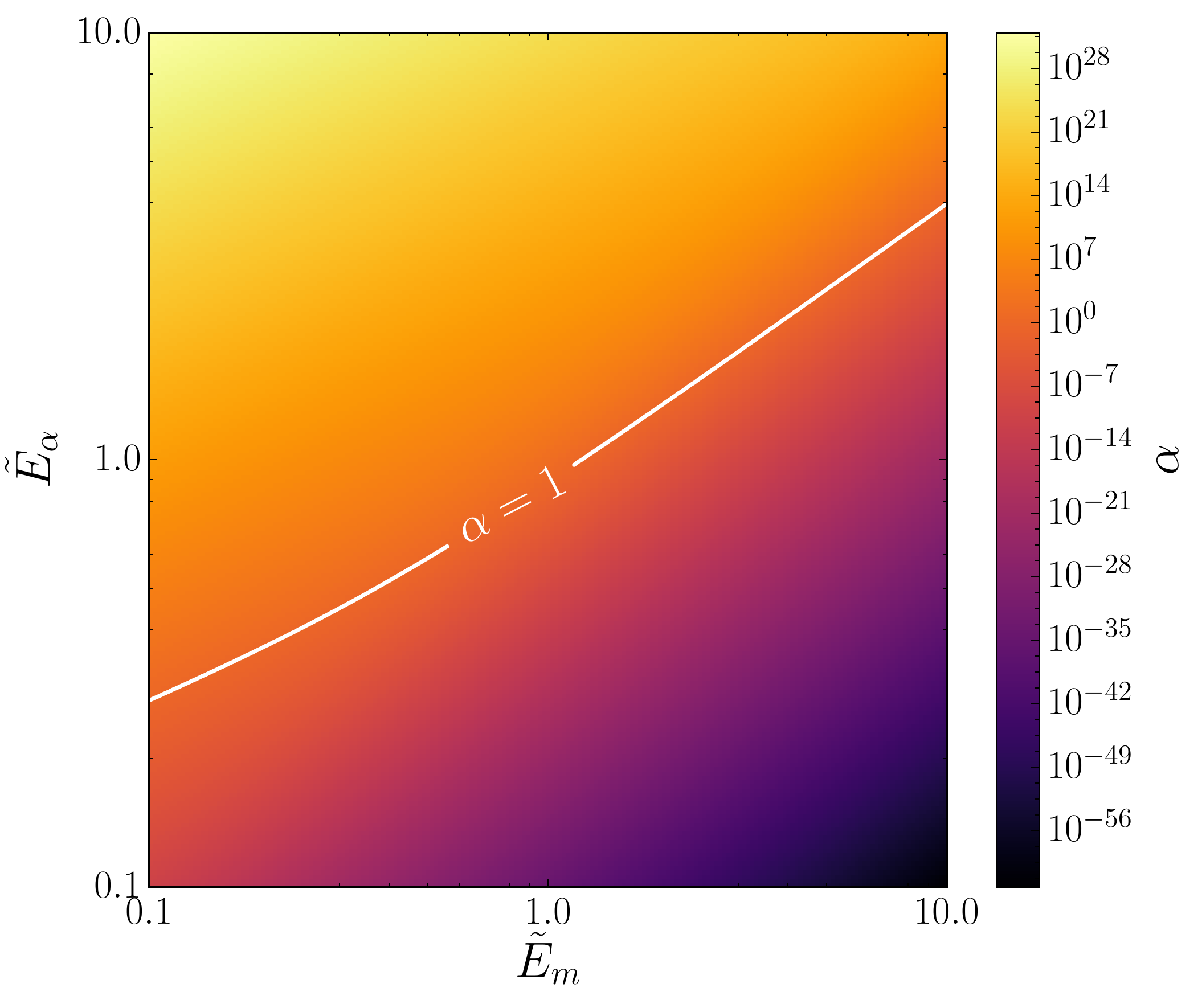}
 \caption{\it Values of $\alpha$ as a function of $\tilde{E}_m$ and 
$\tilde{E}_\alpha$ in the range $[0.1, 10.0]$.}\label{fig:alphas}
\end{figure}

\begin{figure*}[!t]
 \includegraphics[width=\columnwidth]{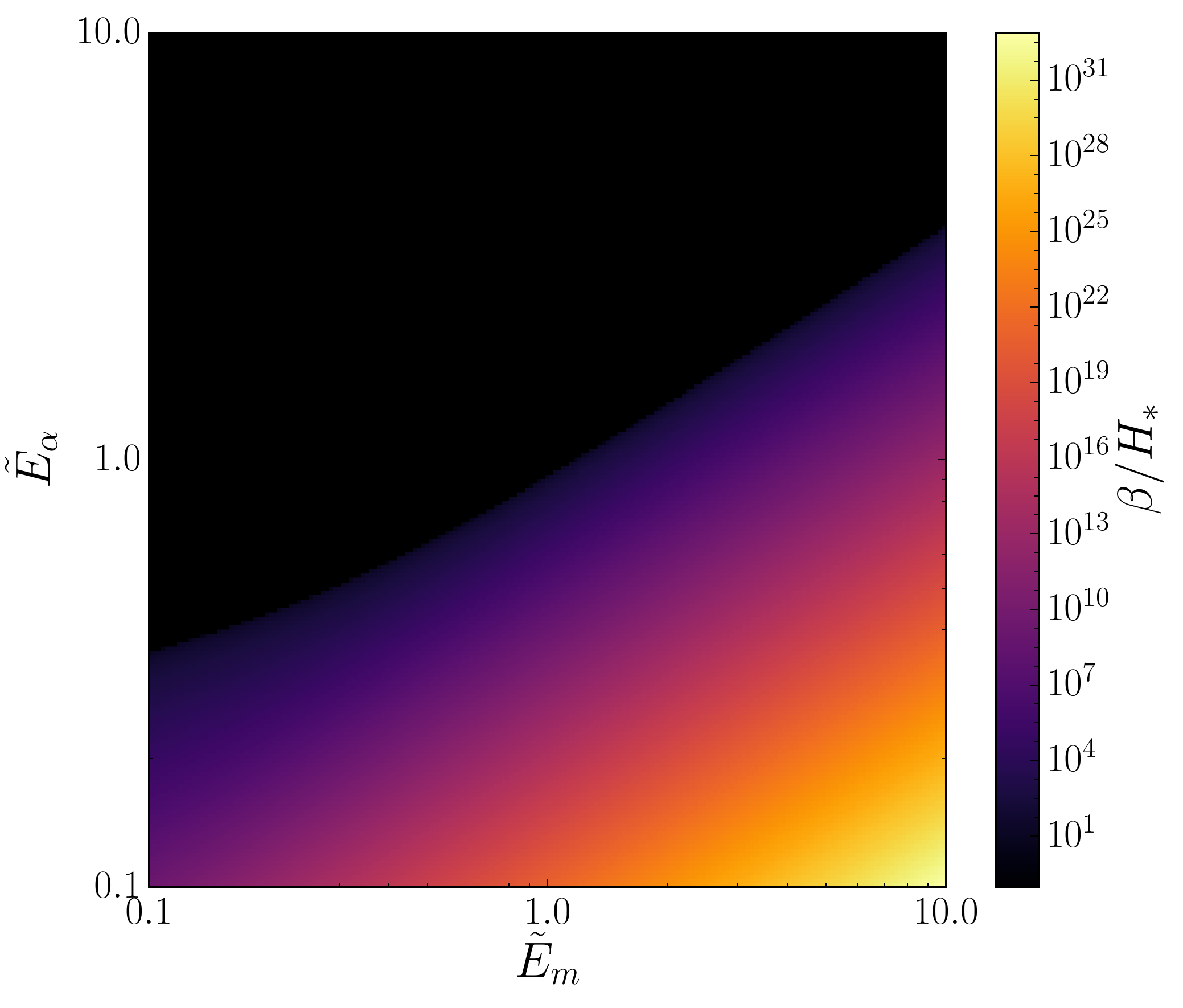}
 \includegraphics[width=\columnwidth]{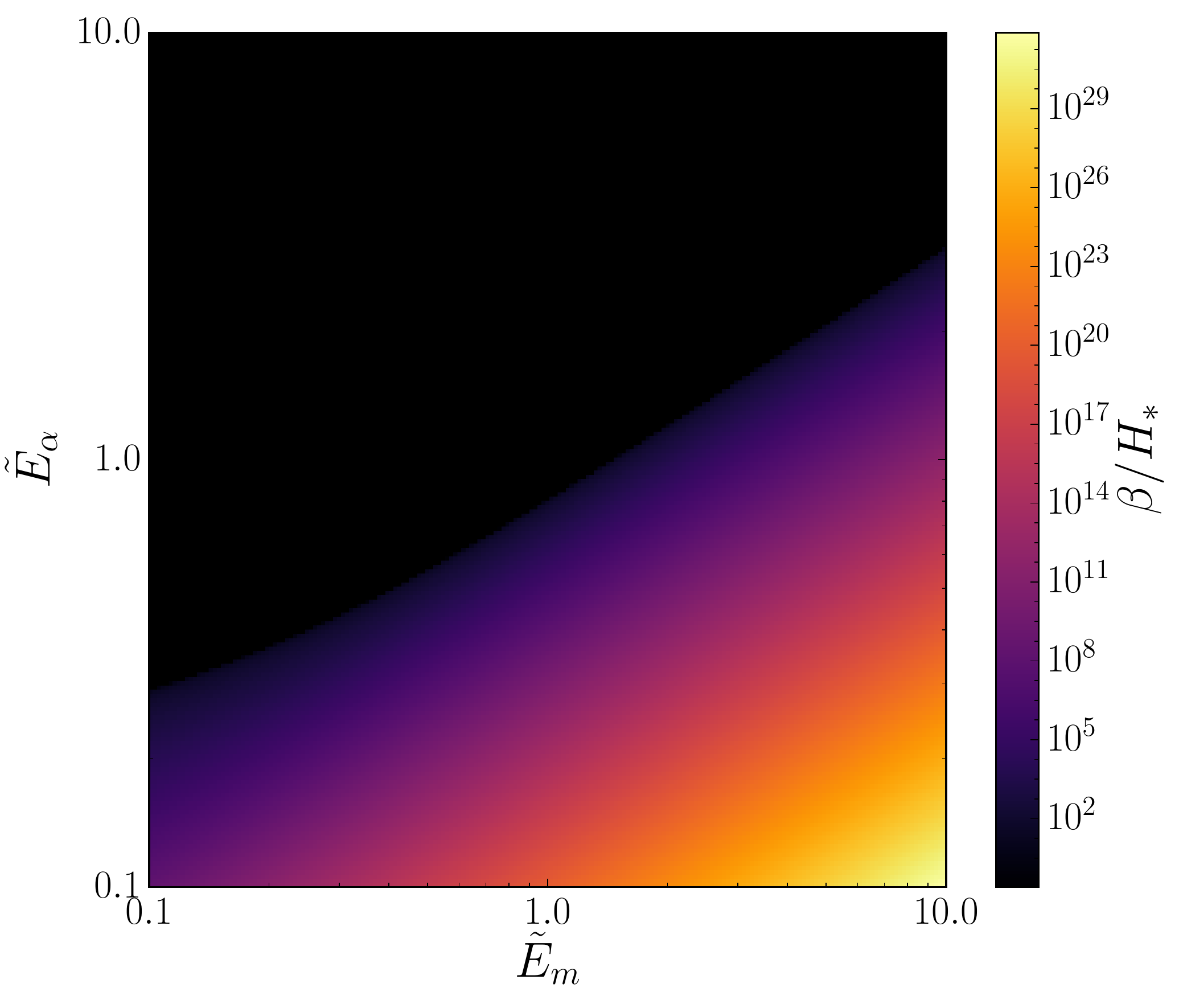}
 \includegraphics[width=\columnwidth]{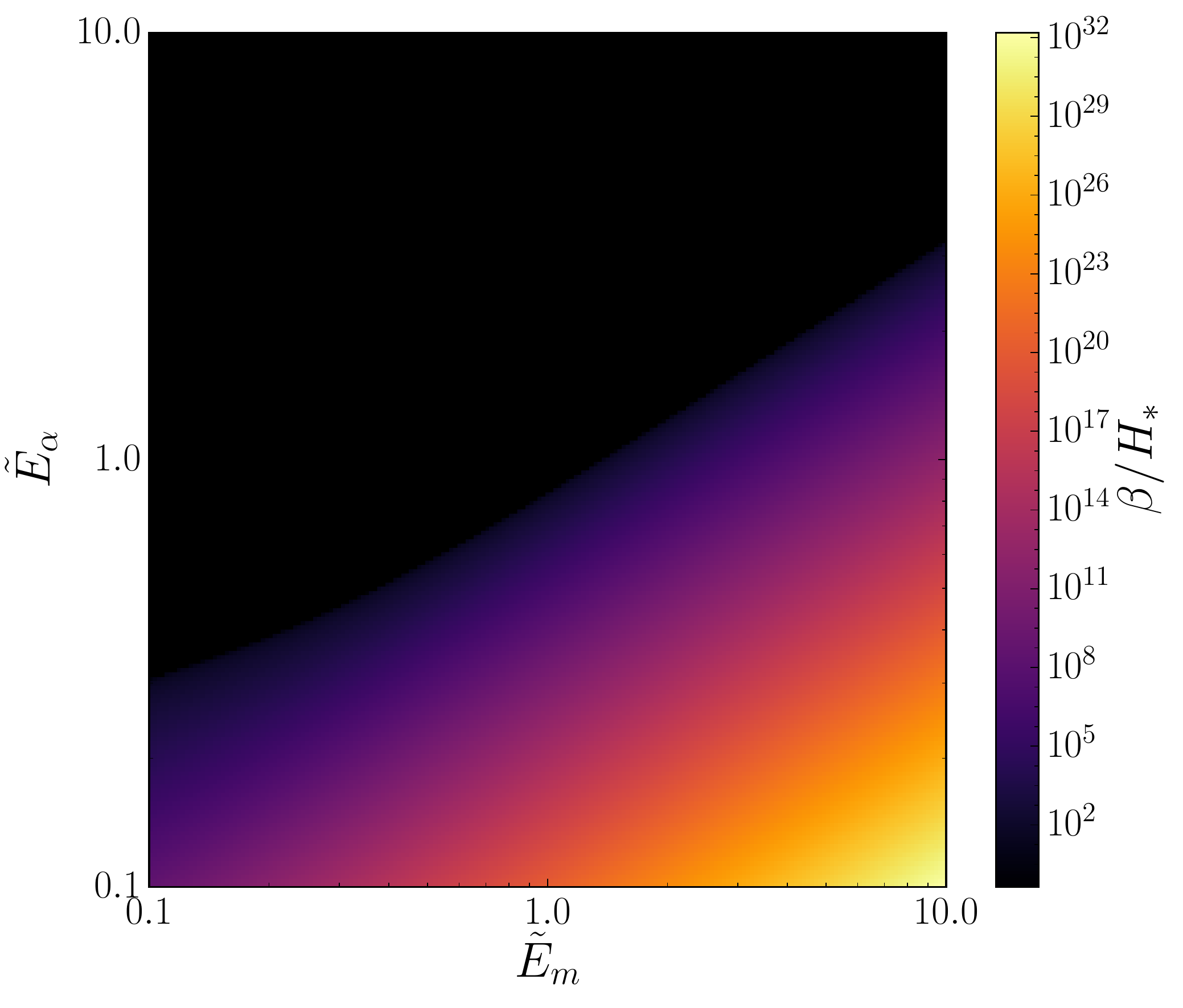}
 \includegraphics[width=\columnwidth]{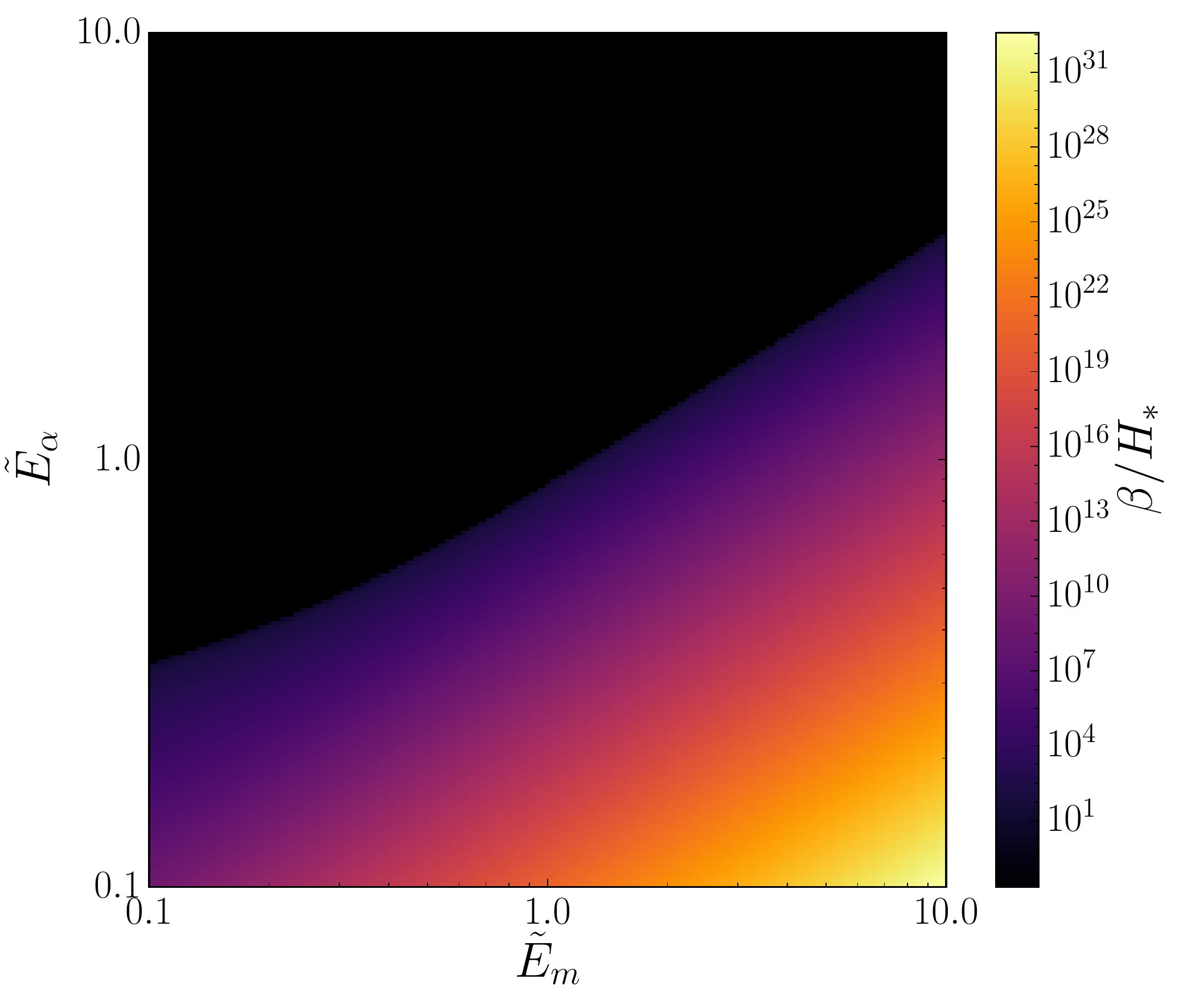}
 \caption{\it Values of $\beta/H_*$ as a function of $\tilde{E}_m$ and 
$\tilde{E}_\alpha$ in the range $[0.1, 10.0]$ for different values of $a_T$. From 
top to bottom and left to right we show $a_T=1, 0.1, 0.2$ and 
$0.5$.}\label{fig:betas}
\end{figure*}

The validity of the high-temperature approximation assumed in Eq.~\eqref{VTa} 
is easy to check. It is equivalent
to requiring ${T}_* >v$, namely %which using Eq.~\eqref{eq:Tc} results in
\begin{equation}
%
 %{\tilde{S}_3^{\,\rm cl}} >100\,.
 \tilde{T}_* > 1\,.
 \label{eq:Tchigh}
\end{equation}

Thus, for any shape of the effective potential at $\tilde{T}_*$, e.g.
that plotted in Fig.~\ref{fig:pot2},
the expression for the effective potential at general $\tilde{T}$ in \eqref{VTa} is justified when Eq.~\eqref{eq:Tchigh} holds.

In summary, to obtain $\beta/H_*$, we need to find the bounce solution in the 
theory with the effective potential (we now use the
dimensionless variables),
\begin{equation}
\tilde{V}_{\tilde{T}} (\tilde{\varphi})\,=\, V_{\tilde{T}_*} (\tilde{\varphi}) 
\,+\, a_T\, (\tilde{T}^2-\tilde{T}_*^2)\, \tilde{\varphi}^2\,,
\label{VTasc}
\end{equation}
compute the 3D action on the bounce, $\tilde{S}_3^{\,\rm cl}(\tilde{T})$, and 
finally evaluate,
\begin{equation}
\frac{\beta}{H_*}\, =\, 
\tilde{T}_*\frac{d}{d\tilde{T}}\left(\frac{1}{\tilde{T}}\,\tilde{S}_3^{\,\rm 
cl}(\tilde{T})\right)_{\tilde{T}=\tilde{T}_*}.
\label{eq:betasc}
\end{equation}

There are three free parameters in total characterising the temperature-dependent potential \eqref{eq:Tchigh} and hence
the classical action: $\tilde{E}_\alpha$,  $\tilde{E}_m$ and $a_T$. From these 
we obtain the three key parameters for the
GW spectrum: the nucleation temperature $T_*/v$, the latent heat $\alpha$ and the $\beta$ parameter using 
Eqs.~\eqref{eq:Tc}, \eqref{eq:alph}, \eqref{VTasc} and \eqref{eq:betasc}.

The values of $T_*/v$ in the plane $(\tilde{E}_m, \tilde{E}_\alpha)$ are 
plotted in Fig.~\ref{fig:temperatures}. The region where $\tilde{T}_* = 1$ is 
also shown by the white solid line. Below this line, $\tilde{T}_* > 1$, 
therefore $T_* > v$ and hence the high-temperature approximation for computing 
$\beta/H_*$ as written in Eq.~\ref{eq:betasc} holds.

We show the values of $\alpha$ in the same plane in Fig.~\ref{fig:alphas}. 
Analogously, in Fig.~\ref{fig:betas} we depict the values of $\beta/H_*$ in the 
same plane for four different choices of $a_T$. From top to bottom and left to 
right we have $a_T= 1,\, 0.1,\, 0.2$ and $0.5$. In the black area, the 
high-temperature 
approximation fails. If in a specific model the temperature 
dependence is not quadratic, then $\beta/H_*$ cannot be estimated from the 
plots. Still, in such case one could consider the family of potentials parameterised by $\tilde{T}$ (equivalent to Eq.~\eqref{VTasc}), then for each case read $\tilde{S}_3^\text{cl}$ from Fig.~\ref{fig:temperatures} (roughly $\tilde{S}_3^\text{cl}\sim 140\,\tilde{T}_*$), and then take the corresponding derivative to compute $\beta/H_*$.
%but of course our procedure would still work.

\begin{figure}[t]
 \includegraphics[width=\columnwidth]{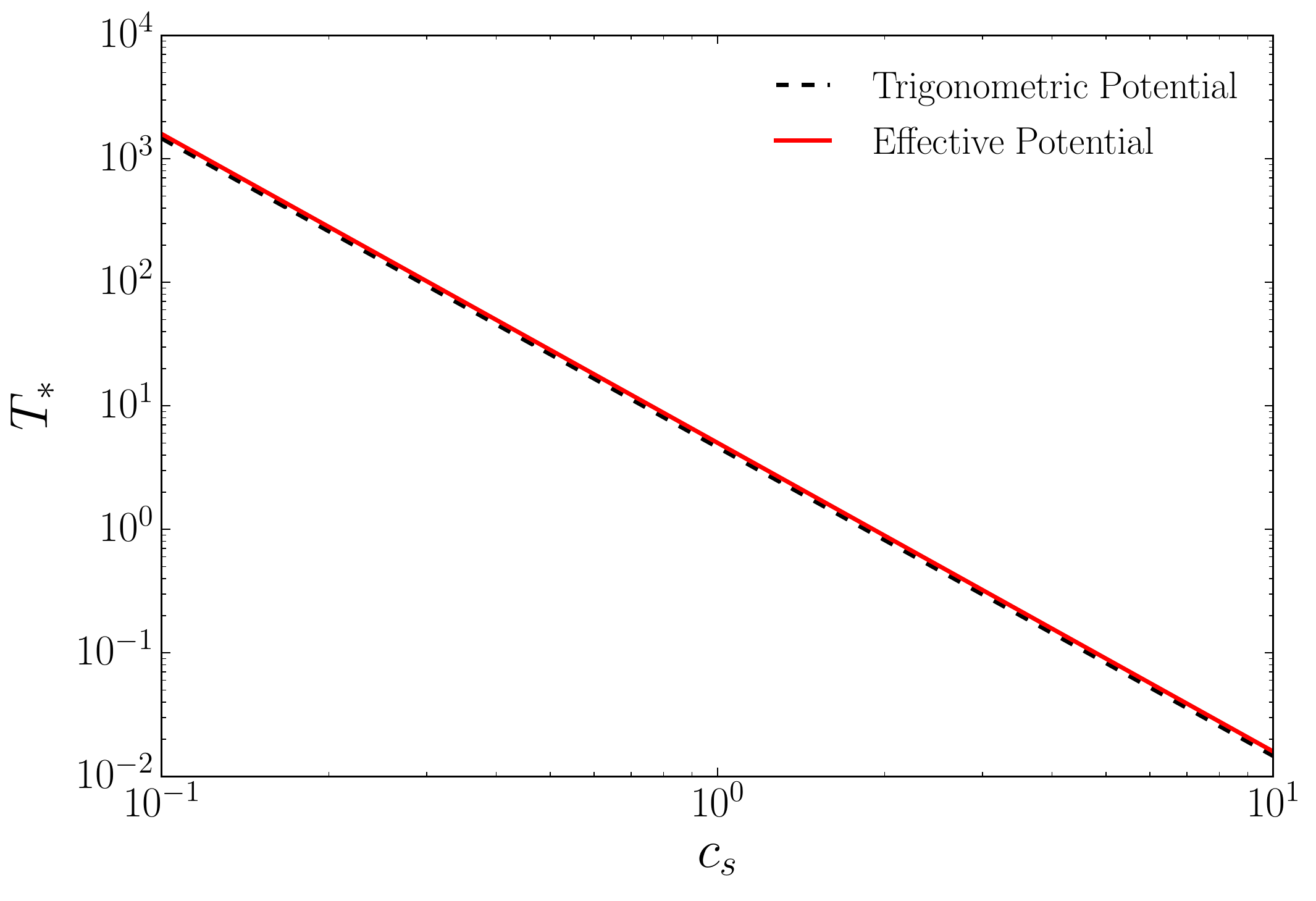}
 \caption{\it Value of $T_*$ as a function of $c_s$ in the 
model defined by Eq.~\ref{eq:sin} computed using the effective potential
(solid red) and solving the bounce equation from scratch 
(dashed black).}\label{fig:sin}
\end{figure}

\begin{figure*}[!t]
 \includegraphics[width=\columnwidth]{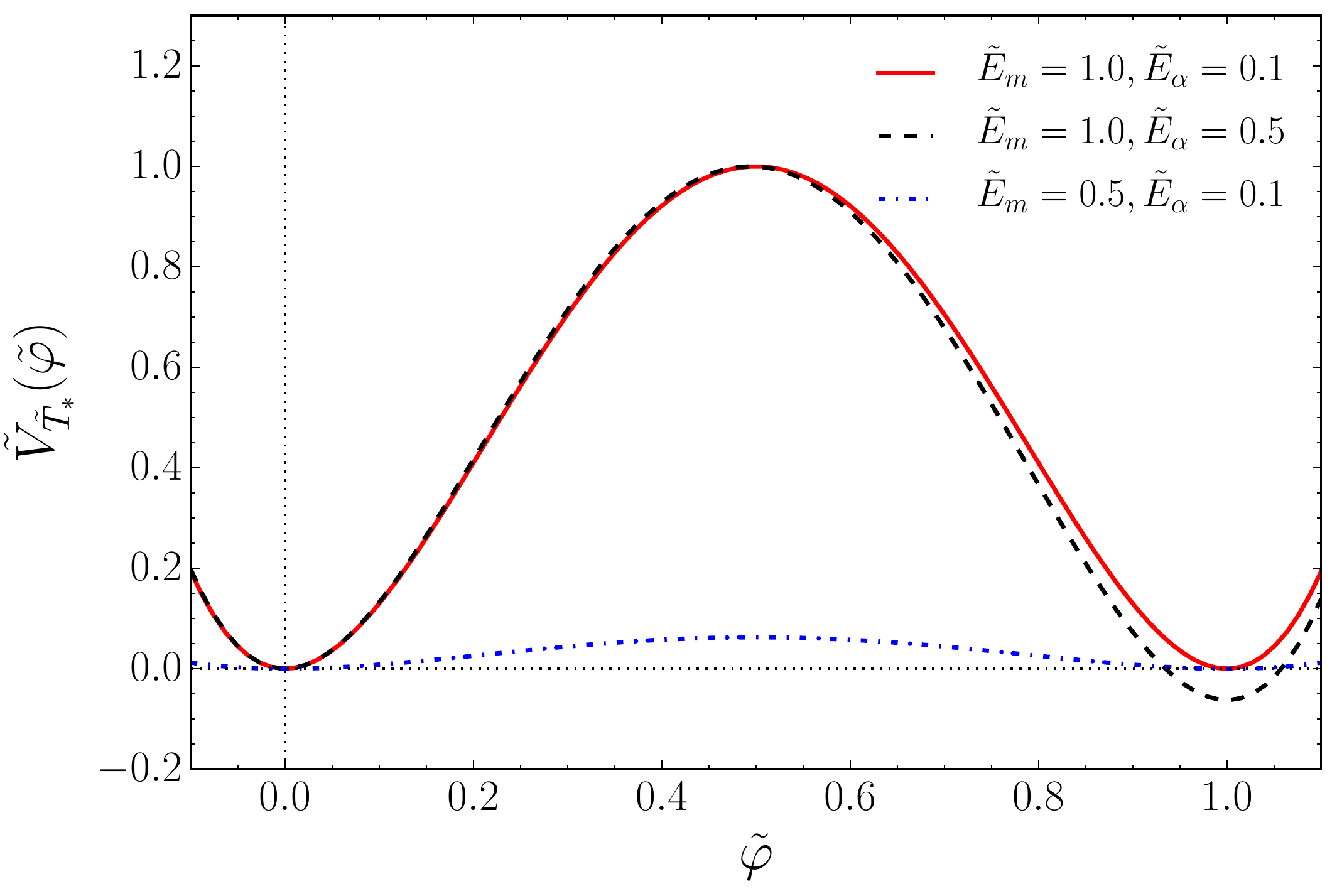}
 \includegraphics[width=\columnwidth]{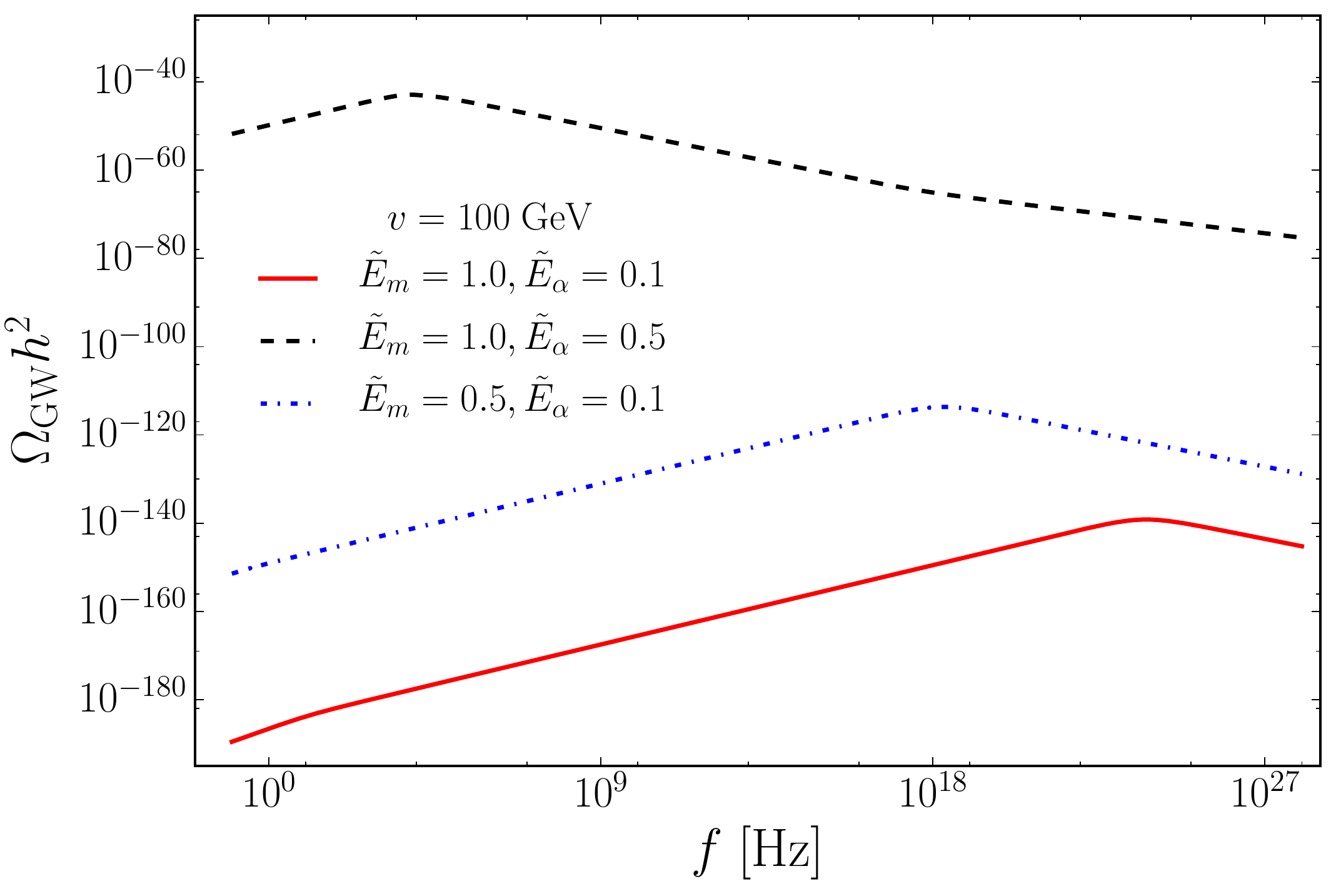}
 \caption{\it In the left panel we show the scalar potential for $\tilde{E}_m = 
1, \tilde{E}_\alpha = 0.1$ (solid red); $\tilde{E}_m = 1, 
\tilde{E}_\alpha = 0.5$ (dashed black) and $\tilde{E}_m = 0.5, 
\tilde{E}_\alpha = 0.1$ (dashed-dotted blue) at $\tilde{T}_*$. 
(Note that the height of the barrier and the relative vacuum energy density depend on the fourth power of the parameters
$\tilde{E}_m$ and $\tilde{E}_\alpha$.)
The right 
panel shows the  
stochastic spectra of GWs produced in the course of the FOPTs for these choices of the scalar potential for $v=100$ GeV.}\label{fig:comparison}
\end{figure*}

Finally, we test the robustness of our parametrisation of the potential by computing $T_*$ in a 
highly non-polynomial potential given by
\begin{equation}\label{eq:sin}
 V_T(h) =  h \sin{(c_s h)}~,
\end{equation}
and comparing it to the result obtained by just 
plugging the values of $E_m$, $E_\alpha$ and $v$ (which depend on $c_s$) extracted from the expression \eqref{eq:sin} 
 into our parametrisation 
in Eq.~\ref{eq:parametrisation}. The results are shown in Fig.~\ref{fig:sin}. 
Notably, our method provides a reasonable estimate of the nucleation 
temperature also in this case, demonstrating that $E_m$ and 
$E_\alpha$ are the main global characteristics of the scalar potential.
We have also checked that even nearly conformal potentials can 
be well described by these global characteristics. For example, we have studied 
the potential of a meson-like dilaton~\cite{Bruggisser:2018mrt}. Disregarding 
its mixing with the Higgs, it reads parametrically:
\begin{equation}
 V(\chi)\sim a_\chi\chi^4 - \epsilon(\chi)\chi^4~,
\end{equation}
with $\epsilon(\chi)\sim 
b_\chi(\chi/\chi_0)^\gamma/[1-c_\chi(\chi/\chi_0)^\gamma]$ the conformal 
breaking function and $a_\chi$, $b_\chi$, $c_\chi$, $\chi_0$ and $\gamma$ 
constants. For the case represented in the right panel of Fig. 6 in that 
reference, we have computed $\tilde{E}_m$ and $\tilde{E}_\alpha$, obtaining 
$\sim 0.18$ and $\sim 0.4$, respectively. Within our approach, this gives an 
action $S_3\sim 10066$ GeV. The authors of Ref.~\cite{Bruggisser:2018mrt} use 
$S_3/T_* = 140$ as the criteria to obtain $T_*$. Using the same criteria, we 
obtain therefore $T_*\sim 71.9$ GeV, while they report $T_*\sim 65.6$ GeV. 
This implies an error smaller than 10 \%.

\section{Calculating the stochastic gravitational wave spectrum}
\label{sec:gw}
Following Ref.~\cite{Caprini:2015zlo}, we estimate the stochastic GW background 
as the linear 
combination of three pieces:
\begin{equation}\label{eq:gw}
 h^2\Omega_\text{GW}\sim h^2\Omega_\varphi + h^2\Omega_\text{sw} + 
h^2\Omega_\text{turb}~.
\end{equation}
The first component describes the contribution of the field $\varphi$ itself, 
due to the collisions of bubble walls after nucleation. Numerical 
simulations~\cite{Huber:2008hg} suggest that it is approximately given by
%
%\begin{widetext}
\begin{align}
 h^2\Omega_\varphi &\sim 1.67\times 10^{-5}\\\nonumber
 &\times\mathcal{F}(2,2)
 \bigg(\frac{0.11 
v_w^3}{0.42+v_w^2}\bigg)\bigg[\frac{3.8\, 
(f/f_\text{env})^{2.8}}{1+2.8\,(f/f_\text{env})^{3.8}}\bigg]~,
\end{align}
%\end{widetext}
%
with
\begin{equation}
\mathcal{F}(x,y) = \bigg(\frac{H_*}{\beta}\bigg)^x\bigg(\frac{\kappa\alpha}{
1+\alpha } \bigg)^y~ 
\bigg(\frac{100}{g_*}\bigg)^{1/3}~,
\end{equation}
$f_\text{env} \sim 16.5\times 10^{-3}\, \text{mHz}\, 
[0.62/(1.8-0.1v_w+v_w^2)]\, \mathcal{C}$, and $\mathcal{C}$ given by
\begin{equation}
 \mathcal{C} = \bigg(\frac{\beta}{H_*}
\bigg)\bigg(\frac{T_*}{100\,\text{GeV}}\bigg)\bigg(\frac{g_*}{100}\bigg)^{1/6}~.
\end{equation}
We remind that $H_*$ and $g_*$ stand for the 
Hubble parameter and the number 
of relativistic degrees of freedom in the plasma at $T=T_*$, respectively. (Hereafter we will restrict to the regime where $\alpha\lesssim 1$, to avoid significant reheating, so that the temperature after the FOTP completes is indeed $\sim T_*$.)
$v_w$ represents the bubble wall velocity and $\kappa$ the fraction of latent 
heat transformed into kinetic energy of $\varphi$. 

The second term in Eq.~\ref{eq:gw} represents the GW background due to sound 
waves produced after the collision of bubbles and before the 
expansion dissipates the kinetic energy in the plasma. It comprises the dominant source of GW radiation. It approximately reads
%
%\begin{widetext}
\begin{align}
 &h^2\Omega_\text{sw} \sim 
2.65\times 
10^{-6}\\\nonumber
&\times\mathcal{F}(1,2)\, v_w\, 
\bigg(\frac{f}{f_\text{sw}}\bigg)^3\bigg(\frac{7}{4+3\,(f/f_\text{sw})^2}\bigg)^
{ 7/2 }~,
\end{align}
%\end{widetext}
%
with $f_\text{sw} \sim 1.9\times 10^{-2} 
\,\text{mHz}\,(1/v_w)\, \mathcal{C}$.
%
%  %
 %%
%

Finally, $h^2\Omega_\text{turb}$ is the magnetohydrodynamic turbulence formed 
in the plasma after the collision of bubbles:
%
%\begin{widetext}
 \begin{align}
  &h^2\Omega_\text{turb} \sim 3.35\times 
10^{-4}\\\nonumber
&\times\mathcal{F}(1,3/2) v_w \bigg\lbrace 
\frac{(f/f_\text{turb})^3}{[1+(f/f_\text{turb})]^{11/3} (1+8\pi 
f/h_*)}\bigg\rbrace~,
 \end{align}
%\end{widetext}
%
%
%
with 
%
% \begin{widetext}
%  %
%  \begin{equation}
  %
 $f_\text{turb} \sim 2.7\times 
10^{-2}\,\text{mHz}\,(1/v_w)\,\mathcal{C}$  and $h_*$ being the redshifted 
Hubble 
time, $h_* 
= 16.5\times 10^{-3}\,\text{mHz}\,T_* (g_*/100)^{1/6}/(100\,\text{GeV})$.

The bubble wall velocity $v_w$ is hard to estimate in general. It has been shown however that if the runaway condition $\alpha > \alpha_\infty \sim 4.9\times 10^{-3}/\tilde{T}_*^2$ is satisfied, then the bubble wall velocity is likely $v_w\sim 1$~\cite{Espinosa:2010hh,Caprini:2015zlo}. This happens in most of our parameter space. Moreover, the GW spectrum does not change dramatically in the allowed range of $v_w$ (conservative estimates suggest that $v_{w,\text{min}}> 1/\sqrt{3}>0.5$~\cite{Steinhardt:1981ct}), so we fix $v_w\sim 1$ for simplicity.
For $\kappa$, we take the fit~\cite{Espinosa:2010hh,Caprini:2015zlo}

\begin{equation}
 \kappa\sim \alpha (0.73+0.083 \sqrt{\alpha} + \alpha)^{-1}\,.
\end{equation}
In Fig.~\ref{fig:comparison}, we show the GW stochastic background 
corresponding to different shapes of the potential at the nucleation 
temperature. We note that, for a barrier of fixed height, increasing the depth of the true 
vacuum shifts the spectrum to smaller frequencies (because it reduces $T_*$; see also 
Fig.~\ref{fig:temperatures}) while it enhances the amplitude of the GW 
spectrum. 
The GW signal is also shifted to smaller frequencies and enhanced in
amplitude 
if the barrier is decreased for a fixed value of the depth of the true 
vacuum, although the effect in this direction is less pronounced.

Clearly, reconstruction of the GW spectrum at future facilities (see 
Refs.~\cite{Figueroa:2018xtu} for ongoing works at LISA) could shed light on the 
global 
properties of the scalar potential.

% \
% T

\section{Limits from present and future gravitational wave experiments}
\label{sec:ligo}
\begin{figure}[t]
 \includegraphics[width=\columnwidth]{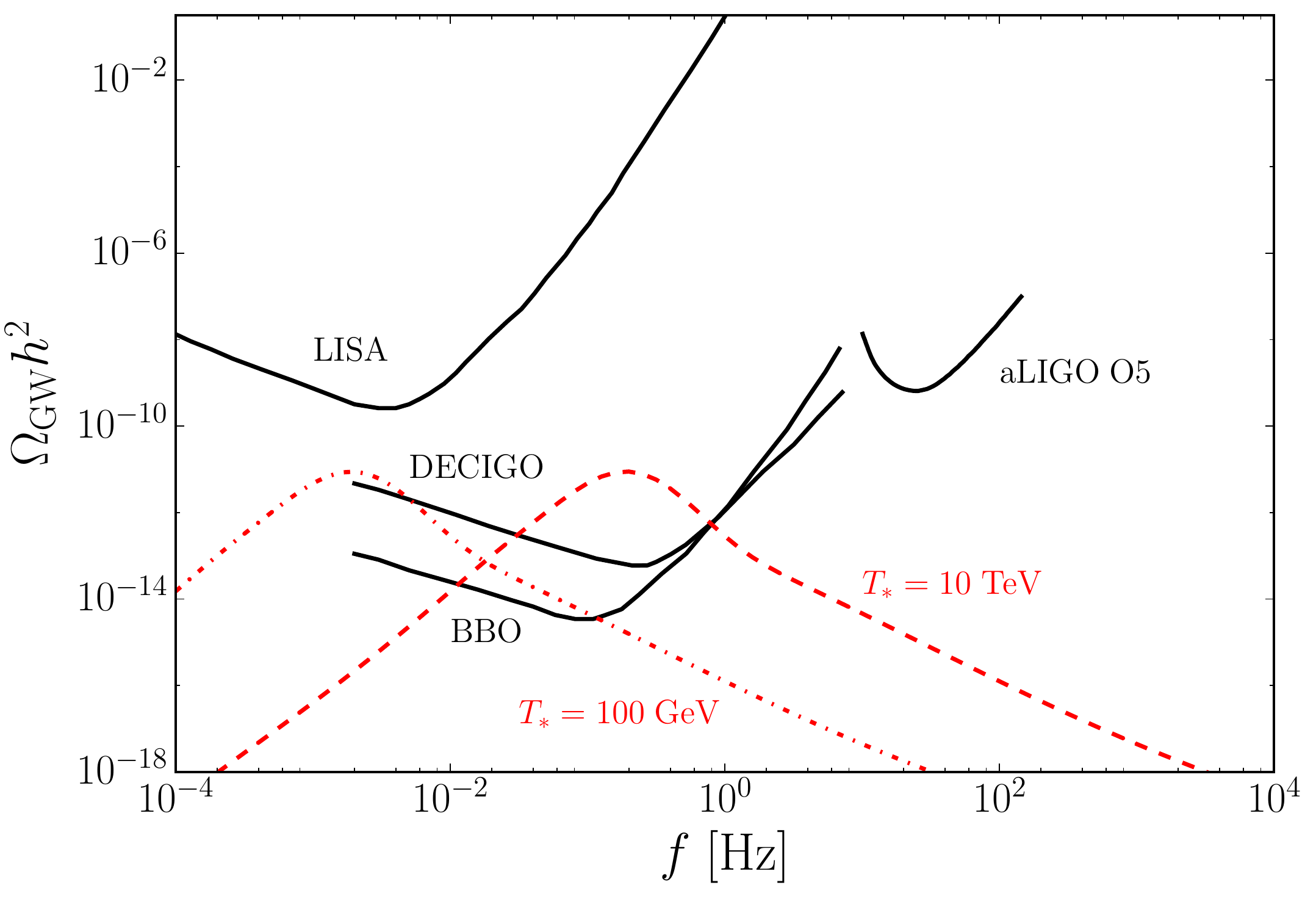}
 \caption{\it Sensitivity curves of LISA, DECIGO, BBO and 
aLIGO O5 to GWs as a function of the frequency. The 
predicted GW stochastic backgrounds for $\alpha =0.1$ and 
$\beta/H_* = 100$ for $T_* = 100$ GeV (dashed-dotted red 
curve) and $T_* = 10$ TeV (dashed red curve) are
also shown for comparison.}\label{fig:sensitivity}
\end{figure}

The sensitivity curves of different GW observatories are represented in 
Fig.~\ref{fig:sensitivity}. They are taken from the GW plotter~\url{http://rhcole.com/apps/GWplotter}~\cite{Moore:2014lga}, which is based on Ref.~\cite{Sathyaprakash:2009xs} (for LISA~\cite{PPA}) and Ref.~\cite{Yagi:2011wg} (for BBO~\cite{Harry:2006fi} and DECIGO~\cite{Kawamura:2006up}). The GW 
stochastic background for $\alpha = 0.1$ and
$\beta/H_* = 100$ is also plotted for comparison for $T_* = 100$ GeV (dashed-dotted red curve) and $T_* = 10$ TeV (dashed red curve).
In order to address the reach of each of these facilities to the 
GW background originating from a FOPT with an effective potential at $T_*$ 
characterised by $E_\alpha$ and $E_m$, we proceed as follows. For each value of 
$v$ 
%in the range $\sim$ 1 -- $10^{14}$ GeV, we compute 
in the range $\sim$ 1 -- $10^{6}$ GeV, we compute 
$E_\alpha$ and $E_m$ restricting to the region of $\tilde{E}_{m},\tilde{E}_\alpha\in 
[0.1,10]$ where $\alpha\lesssim 1$. We subsequently obtain, from the results 
above, the values of $\alpha$ and $\beta/H_*$ for $a_T = 1$. We finally 
compare the GW spectrum as given by Eq.~\ref{eq:gw} with the sensitivity curves 
depicted in Fig.~\ref{fig:sensitivity}. We naively assume that, if the two 
curves overlap at any point in a fixed experiment, the latter can 
test the corresponding potential\footnote{Using this procedure, we have 
estimated the region in the $(\alpha, \beta/H_*)$ plane that can be tested 
with LISA for $T_* = 100$ GeV and compared it with that given in 
Ref.~\cite{Caprini:2015zlo}. Our results turn out to be slightly more conservative.}. 
Thus, for example, the 
GW spectrum represented by the dashed red curve in Fig.~\ref{fig:sensitivity}
would be observable by DECIGO and BBO but not by LISA or aLIGO O5. 
Let us also emphasize that we are neglecting 
the possible effects 
of  having not ``long-lasting'' sound waves~\cite{Ellis:2018mja,Ellis:2019oqb}. 

\begin{figure}[t]
 \includegraphics[width=\columnwidth]{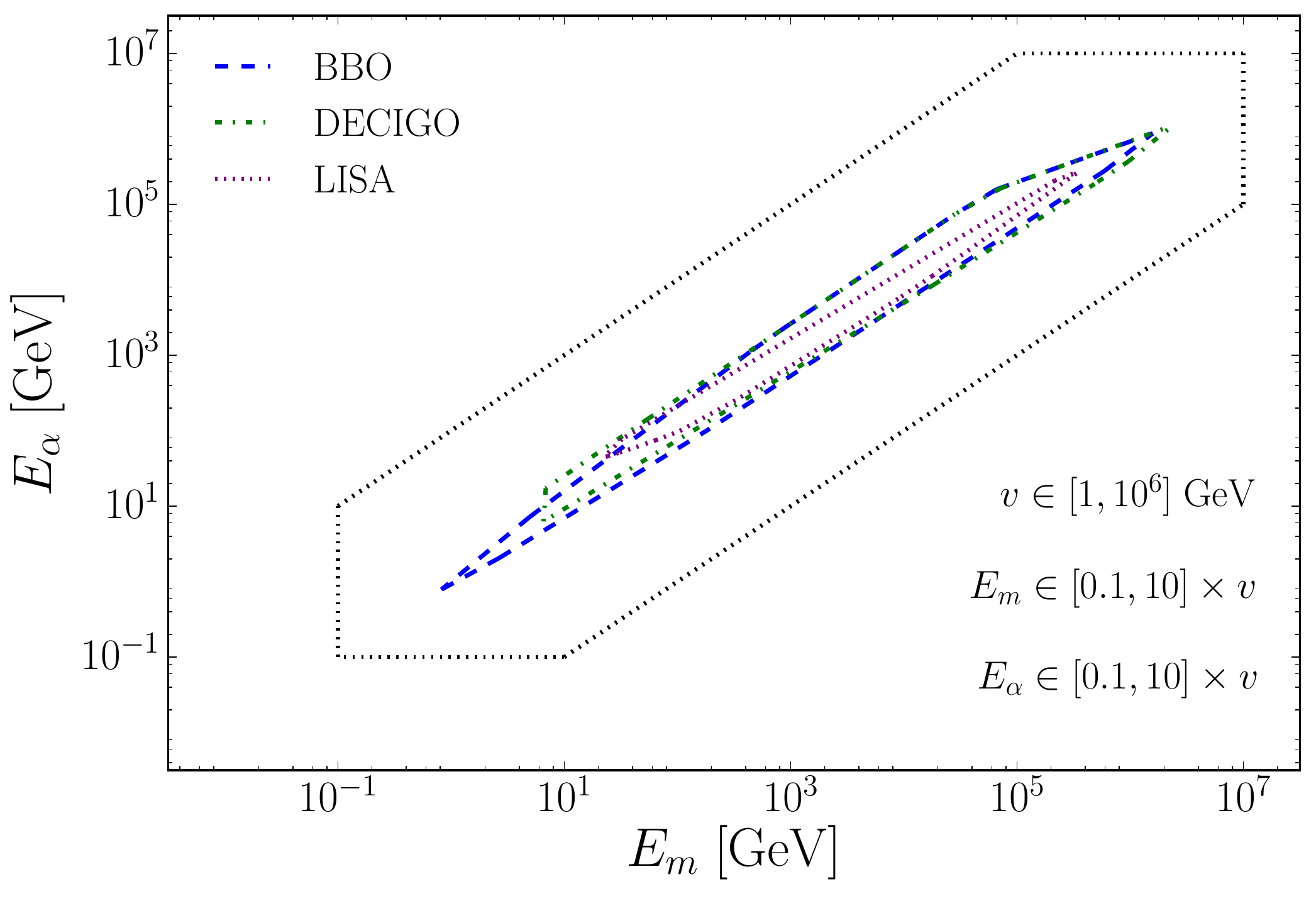}
 \caption{\it Regions of the plane $(E_m, E_\alpha)$ that can 
be probed by  BBO (dashed blue), DECIGO 
(dashed-dotted green) and LISA (dotted purple) for $v$ 
in the range $[1,10^{14}]$ GeV. The dotted black box shows the region of the plane covered by the parameter scan.}\label{fig:reach}
\end{figure}

The results are shown in Fig.~\ref{fig:reach}. We note that %all facilities
BBO, DECIGO and LISA are 
mostly sensitive to the region $E_\alpha\sim (0.1-10)\times E_m$; the 
variations in magnitude between the different experiments being due to their 
different frequency reach. aLIGO O5 is sensitive to similar values, provided $v\gtrsim 10^7$ GeV, which is beyond the regime of applicability of Eq.~\eqref{eq:Tc}.

To obtain the results displayed in this figure, we have scanned over the 
parameter ranges $E_\alpha \in [0.1, 10]\times v$ and $E_m \in [0.1, 10]\times 
v$ simultaneously. This region is depicted in Fig.~\ref{fig:reach} with the 
dotted black box. Thus, moving along the ellipsoid shape in 
Fig.~\ref{fig:reach} to larger and larger values of $E_m$ and $E_\alpha$\footnote{This is a direction of travel over many orders of 
magnitude starting from
 $E_m \sim E_\alpha\sim 10^{-1}$~GeV and reaching to $E_m \sim E_\alpha\sim 
10^{7}$~GeV in Fig.~\ref{fig:reach}.}, implies
 increasing the values of $v$.
In other words, a larger value of $v$ has to be 
compensated by a deeper well of the potential to retain sensitivity at 
GW experiments. The region where $E_\alpha$ is small, and the 
potential barrier, expressed by $E_m$, is large, i.e. the lower right region of Fig.~\ref{fig:reach},
becomes experimentally 
inaccessible. 

The 
upper left region of Fig.~\ref{fig:reach}, where the potential well is deep, i.e.
$E_\alpha$ is large, and the barrier is small, would result in a very small 
value for the Euclidean action $\tilde{S}_3^{\,\rm cl}$ and, thus, according to 
Eq.~\ref{eq:Tc}, a very small nucleation temperature $T_*$. Such small $T_*$ 
would be formally unacceptable as it would invalidate our assumption of the high 
temperature approximation $T\gg v$ 
in Eq.~\ref{VTa}.
But even more importantly, the potential 
$V_T (\varphi)$ we consider is the result of a dynamical process when the plasma 
is cooling. Therefore, in realistic models, one would expect the phase 
transition to happen at temperatures much larger than that corresponding to
the potential with
the parameters in the upper left region of Fig.~\ref{fig:reach}.
Namely in a region closer to the ellipsoid shape in 
Fig.~\ref{fig:reach}.

\section{Connection to fundamental theories}
\label{sec:models}

Our results do not only show the interplay between the global properties of the 
scalar potential and the GW stochastic background; they can be also used to 
compute the latter in an arbitrary model of new physics without necessarily 
solving for the bounce in Eq.~\ref{eq:S3d} from scratch. To this aim, we 
provide tables with precomputed values of $\tilde{T}_*$, $\alpha$ and $\beta/H_*$ for 
varying $\tilde{E}_m$ and $\tilde{E}_\alpha$; see the webpage~\url{https://www.ippp.dur.ac.uk/~mspannow/gravwaves.html}~. Given this:

\begin{enumerate}
 \item For fixed $T$ and $a_T$, one has to compute the finite-temperature 
effective 
potential in the corresponding model.
\item Subsequently, the values of $v$, $E_\alpha$ and $E_m$ are read off the effective potential. 
The values of $\tilde{E}_\alpha$ and $\tilde{E}_m$ can be trivially obtained 
from the former.
\item Next, one loops over all entries in the table with most 
similar $a_T$ provided in 
the link above. The triad 
$(\tilde{E}_\alpha, \tilde{E}_m, \tilde{T}_*)$ from the table closest to the 
triad made out of the two values obtained in point~2 
and $T/v$ should be 
taken.
\item The Euclidean distance between these two triads 
(normalised to the 
module of the latter), $d$, has to be computed.
%smaller than a predefined cutoff $\epsilon \ll 1$, the values of 
%$\tilde{T}_*$, 
%$\alpha$ and $\beta/H_*$ in the table are to be taken. 
%Otherwise, 
%$T$ should be shifted and one should repeat the procedure starting from point 
%1.
%
\item Points $1-4$ are repeated for different values of 
$T$. The value of $T$ for which $d$ is smallest is taken as the estimated 
$T_*$. The estimated values for $\alpha$ and $\beta/H_*$ are those appearing in the row 
with most similar triad in the corresponding table.
\end{enumerate}
We apply this process to a simple model given by
\begin{align}\label{eq:h6}
 V_T(h) = -\frac{1}{2}\mu h^2 + \frac{1}{4}\lambda h^4 + 
\frac{c_6}{8\Lambda^2} h^6
+ a_T T^2 h^2~.
\end{align}
This Lagrangian captures the modification on the Higgs 
potential due to new physics at a scale 
$\Lambda$~\cite{Grojean:2004xa,Delaunay:2007wb,Chala:2018ari}. For every 
$c_6/\Lambda^2$, we compute $\mu$ and $\lambda$ by 
requiring that the Higgs mass and the electroweak VEV match the measured values 
$m_h\sim 125$ GeV and $v_\text{EW}\sim 246$ GeV. We fix $a_T = 1/32 (4 
m_h^2/v_\text{EW}^2+3g^2+g'^2+4y_t^2-12 c_6 v_\text{EW}^2/\Lambda^2)$, with 
$g$ and $g'$ the $SU(2)_L$ and $U(1)_Y$ gauge couplings, respectively, and 
$y_t$ the top Yukawa. 

\begin{figure}[t]
 \includegraphics[width=\columnwidth]{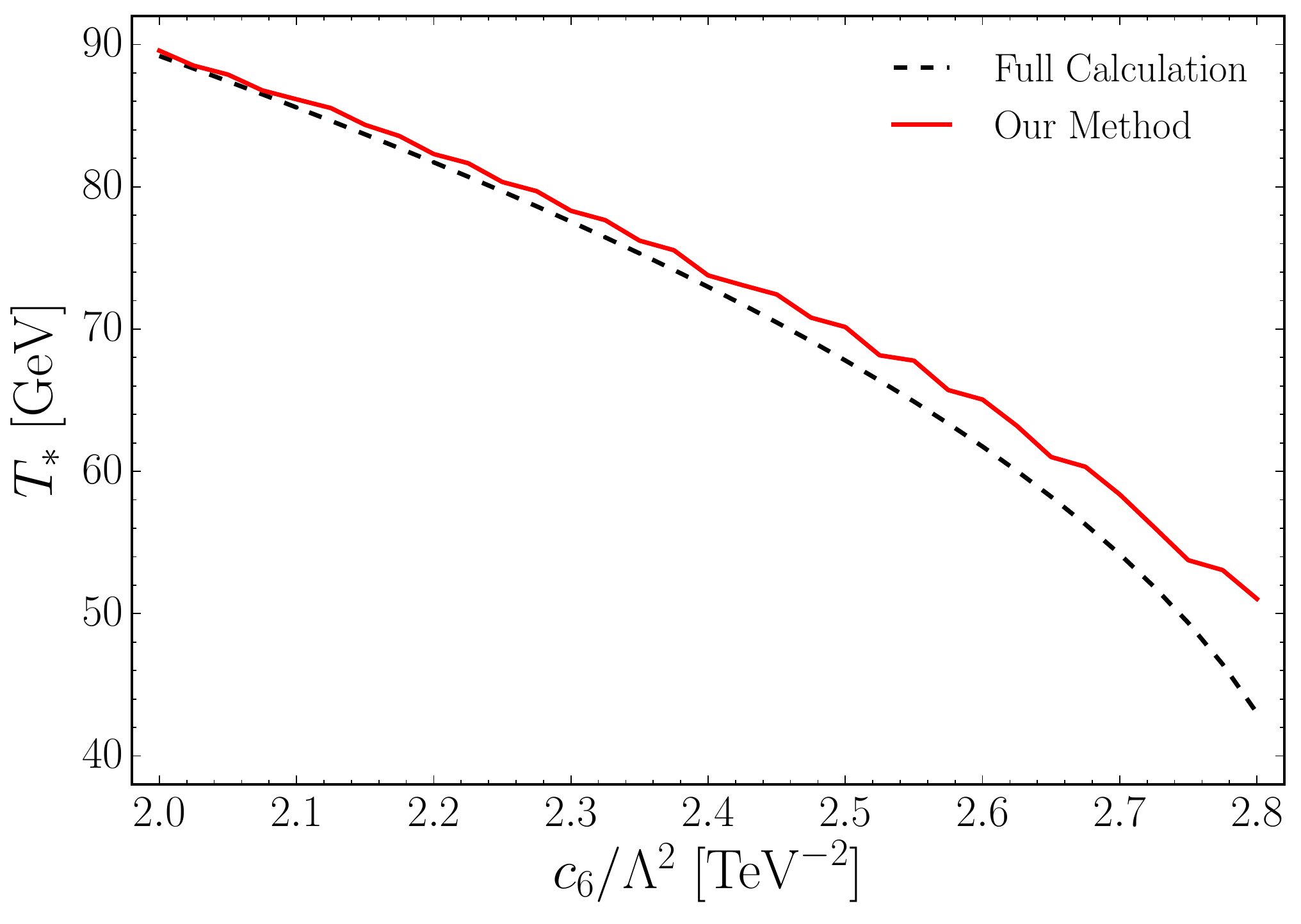}
 \caption{\it Value of $T_*$ as a function of $c_6/\Lambda^2$ in the 
model defined by Eq.~\ref{eq:h6} computed using the method outlined in the text 
(solid red) and solving the bounce equation from scratch 
(dashed black).}\label{fig:h6}
\end{figure}

The value of 
$T_*$ as a 
function of $c_6/\Lambda^2$ obtained using the procedure outlined 
above is shown in Fig.~\ref{fig:h6}. For comparison, we 
also show the value of $T_*$ obtained upon solving the bounce equation with 
\texttt{BubbleProfiler} in this particular model. The goodness of our method 
is apparent.\\

\section{Conclusions and Outlook}
\label{sec:conclusion}
We have computed the GW stochastic background produced in a FOPT triggered by 
the sudden change of VEV of a scalar field with potential characterised by 
given energy barrier $(E_m^4)$ and depth of the true minimum $(E_\alpha^4)$; 
see 
Fig.~\ref{fig:pot1}. We have shown that these parameters capture the most 
important and global characteristics of the scalar potential; the computation 
of the tunnelling rate, nucleation temperature, etc. being highly independent 
of other properties.

We have found that, for fixed values of $E_m$ ($E_\alpha$), the amplitude of the GW spectrum increases for growing (decreasing) $E_\alpha$ ($E_m$), with the frequency peak of the GW spectrum behaving conversely; 
GW observatories being mostly sensitive to the region $E_\alpha\sim 
(0.1-10)\times E_m$.

The reconstruction of the GW stochastic background at future facilities could 
therefore pinpoint the global structure of the Higgs potential, of which we only 
know its shape in a vicinity of the electroweak VEV. (Likewise for other scalar 
fields.)
Thus, this 
study complements previous works in the literature aimed at characterising 
the 
nature of the Higgs potential using e.g. measurements of 
sphaleron 
energies~\cite{Spannowsky:2016ile}. Measurements of double Higgs 
production~\cite{Dolan:2012ac,No:2013wsa,Adhikary:2017jtu} instead can only reveal local properties of the Higgs sector.
For example, the following simple potential
\begin{equation}
 V(h) =  \frac{m_h^2}{2} (h - v)^2 + \frac{m_h^2}{2 v} (h - 
v)^3 + a_4 (h - v)^4~,
\end{equation}
fullfills trivially $V'(v) = 0$, $V''(v) = m_h^2$ and $V'''(v) = 3m_h^2/v$; exactly as in the Standard Model. However, it
has a barrier at zero temperature for $a_4\sim 1/30$. In fact, assuming a $T$ dependence of the form $a_T T^2 h^2$ with $a_T \sim 0.1$, the model undergoes
a FOPT at $T_*\sim 10$ GeV.

Furthermore, as a bonus, we provide a method to use our results to estimate the 
main 
parameters entering the computation of the GW stochastic background, namely the 
nucleation temperature ($T_*$), the ratio of the energy
density released in the phase transition to the energy
density of the radiation bath ($\alpha$) and the inverse duration 
time of the phase transition ($\beta/H_*$). This method allows the user to 
avoid solving the bounce equations from scratch, and therefore it is on a similar 
footing with other dedicated tools such as 
\texttt{CosmoTransitions}~\cite{Wainwright:2011kj} or 
\texttt{BubbleProfiler}~\cite{Athron:2019nbd}.\\

{\emph{Acknowledgements.}}  

\noindent This work is supported in part by an STFC Consolidated grant. MC is 
funded by the Royal Society under the Newton International Fellowship programme. 
MS is funded by the Humboldt Foundation and is grateful to the University of 
Tuebingen for hospitality during the finalisation of parts of this work.
%

%%%%%%%%%%%%%%%%%%%%%%%%%%%%%%%%%%%%%%%%%%%%%%%%%%%%%%%%%%
\bibliography{references.bib}

\end{document}